\documentclass[aps,prl,twocolumn,superscriptaddress]{revtex4-1}

\usepackage[dvips]{graphicx}
\usepackage{color}
\usepackage{bm}
\usepackage{amssymb}

\usepackage[utf8]{inputenc}
\usepackage{amsmath}
\usepackage{tabularx,colortbl}

\begin{document}

\title{Tunable quantum spin liquidity in the 1/6th-filled breathing kagome lattice}

\author{A.~Akbari-Sharbaf}
\affiliation{Institut Quantique and Département de Physique, Université de Sherbrooke, 2500 boul. de l'Université, Sherbrooke (Québec) J1K 2R1 Canada}

\author{R.~Sinclair}
\affiliation{Department of Physics and Astronomy, University of Tennessee, Knoxville, Tennessee, 37996-1200, USA}

\author{A.~Verrier}
\author{D.~Ziat}
\affiliation{Institut Quantique and Département de Physique, Université de Sherbrooke, 2500 boul. de l'Université, Sherbrooke (Québec) J1K 2R1 Canada}

\author{H.~D.~Zhou}
\affiliation{Key laboratory of Artificial Structures and Quantum Control (Ministry of Education), School of Physics and Astronomy, Shanghai JiaoTong University, Shanghai, 200240, China}
\affiliation{Department of Physics and Astronomy, University of Tennessee, Knoxville, Tennessee, 37996-1200, USA}

\author{X. F. Sun}
\affiliation{Department of Physics, Hefei National Laboratory for Physical Sciences at Microscale, and Key Laboratory of Strongly-Coupled Quantum Matter Physics (CAS), University of Science and Technology of China, Hefei, Anhui 230026, People's Republic of China}
\affiliation{Institute of Physical Science and Information Technology, Anhui University, Hefei, Anhui 230601, People's Republic of China}
\affiliation{Collaborative Innovation Center of Advanced Microstructures, Nanjing, Jiangsu 210093, People's Republic of China}

\author{J.~A.~Quilliam}
\email{jeffrey.quilliam@usherbrooke.ca}
\affiliation{Institut Quantique and Département de Physique, Université de Sherbrooke, 2500 boul. de l'Université, Sherbrooke (Québec) J1K 2R1 Canada}

\date{\today}

\begin{abstract}
 
We present measurements on a series of materials, Li$_2$In$_{1-x}$Sc$_x$Mo$_3$O$_8$, that can be described as a 1/6th-filled breathing kagome lattice. Substituting Sc for In generates chemical pressure which alters the breathing parameter non-monotonically. $\mu$SR experiments show that this chemical pressure tunes the system from antiferromagnetic long range order to a quantum spin liquid phase. A strong correlation with the breathing parameter implies that it is the dominant parameter controlling the level of magnetic frustration, with increased kagome symmetry generating the quantum spin liquid phase. Magnetic susceptibility measurements suggest that this is related to distinct types of charge order induced by changes in lattice symmetry, in line with the theory of Chen \emph{et al.} [Phys. Rev. B {\bf 93}, 245134 (2016)]. The specific heat for samples at intermediate Sc concentration and with minimal breathing parameter, show consistency with the predicted $U(1)$ quantum spin liquid. 

 \end{abstract}


\maketitle

One of the most sought after magnetic phases is the quantum spin liquid (QSL), wherein spins form a highly-entangled quantum ground state that supports fractional spin excitations~\cite{Balents2010}. Two main approaches to the discovery of QSL materials have been especially fruitful in recent years: spin-1/2 kagome antiferromagnets~\cite{Yan2011,Mendels2007,Han2012Nature} and triangular-lattice antiferromagnets near a Mott transition~\cite{Motrunich2005,Pratt2011,Yamashita2008, Yamashita2009, Yamashita2011, Itou2008}. However, much remains to be understood about these experimental QSL candidates and some properties remain difficult to reconcile with theory~\cite{Olariu2008,Fu2015,Yamashita2009}. Hence the search for new QSL candidates based on different mechanisms, for example~\cite{Orain2017,Balz2016}, remains a valuable pursuit. In particular, systems in which Hamiltonian parameters can be continuously tuned may provide a prime opportunity to link theoretical models to experimental phenomena. 

In this Letter, we demonstrate that a high degree of tunability can be achieved with the materials, Li$_2$In$_{1-x}$Sc$_x$Mo$_3$O$_8$, that incorporate both spin and charge degrees of freedom. This family of materials consists of a ``breathing'' kagome lattice (BKL) of Mo ions wherein the triangles that point upward are slightly smaller than those that point downward~\cite{Schaffer2017,Clark2013,Orain2017}, with a ``breathing ratio'' $\lambda = d_\nabla / d_\Delta$. In these particular materials the lattice is 1/6th filled, with one unpaired electron for every 3 Mo sites, and its insulating character is ensured by strong next-nearest-neighbor interactions ($V_1$ on up-triangles and $V_2$ on down-triangles)~\cite{Sheckelton2012}.

\begin{figure}
\begin{center}
\includegraphics[width=3.35in,keepaspectratio=true]{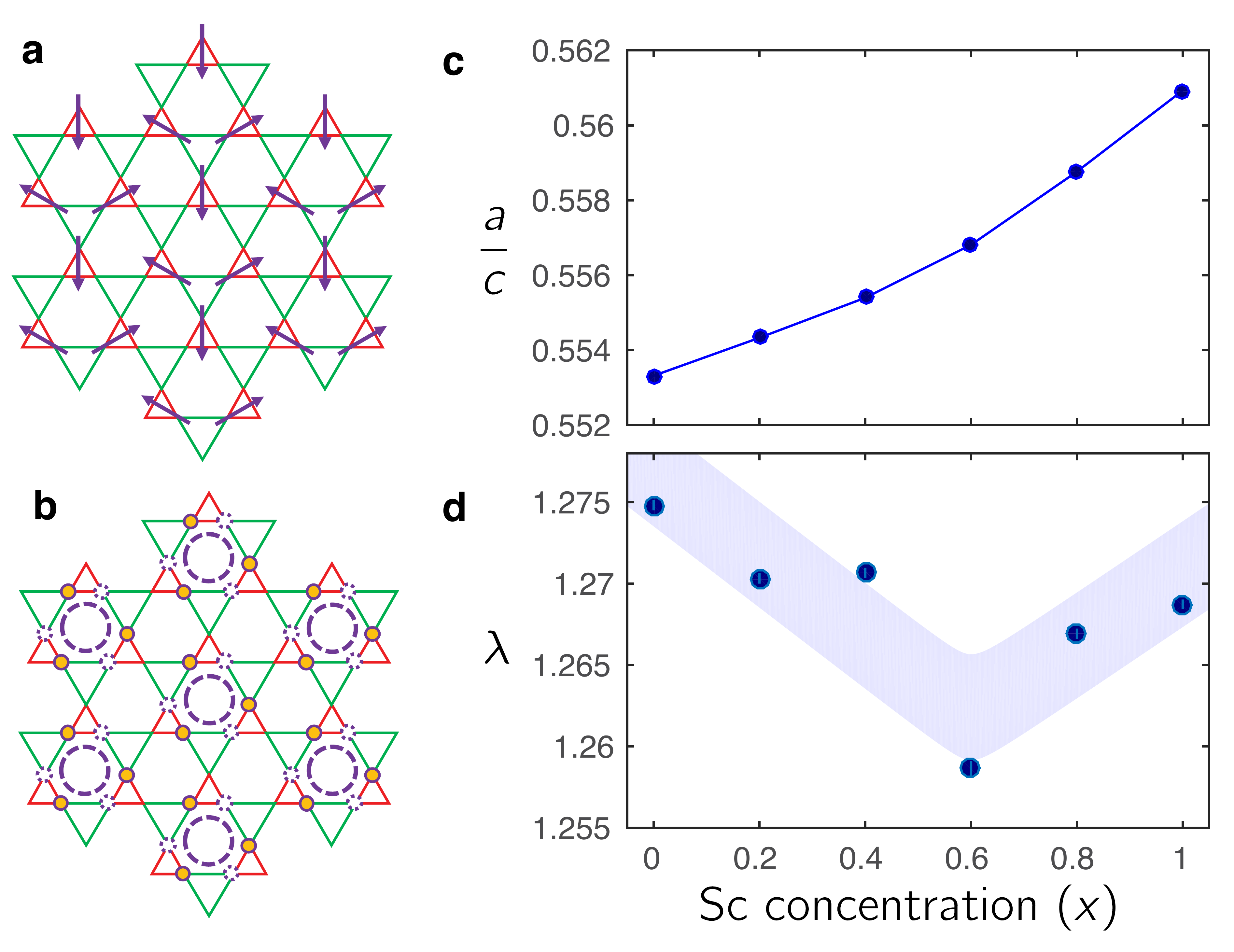}
\caption{Illustrations of (a) the type-I cluster Mott insulator, where electrons are localized on Mo$_3$ units, leading to 120$^{\circ}$ antiferromagnetic order and (b) the PCO state. Resonating hexagons are depicted by dashed circles, and the two spatial configurations of the collective tunnelling electrons are depicted by the open and full circles. (c) The ratio of lattice parameters, $a/c$ and (d) breathing parameter $\lambda$ as a function of $x$. The shaded region is a guide to the eye. Error bars from the refinement are smaller than the data points.} \label{xrd}
\end{center}
\end{figure}

As proposed by Sheckelton \emph{et al.}~\cite{Sheckelton2012} for LiZn$_2$Mo$_3$O$_8$ (LZMO), a similar QSL candidate material~\cite{Mourigal2014}, a plausible charge configuration consists of each electron delocalized over one ``up-triangle'', ultimately leading to a triangular lattice of spin-1/2 moments on Mo$_3$O$_{13}$ clusters, as depicted in Fig.~\ref{xrd}(a). However, it has been pointed out~\cite{Chen2016} that, due to the large spatial extent of the 4$d$ electrons, the single unpaired electrons may have a non-zero probability of tunnelling between adjacent clusters. When $\lambda$ is large, the electrons are expected to localize on the smaller triangles, recovering the Type-I cluster Mott insulator (CMI) proposed by Sheckelton \emph{et al.}~\cite{Sheckelton2012}. When $V_2$ becomes comparable to $V_1$ it is energetically favorable for electrons to collectively tunnel between the small triangles, giving rise to a long range plaquette charge order (PCO), or Type-II CMI, as depicted in Fig.~\ref{xrd}(b). We will show that $x$ in Li$_2$In$_{1-x}$Sc$_x$Mo$_3$O$_8$, tunes the system from a long range ordered (LRO) magnetic phase to a QSL phase and propose that these distinct magnetic phases are a result of the distinct charge configurations. Although the end points of this family (at $x=0$ and $x=1$) have been studied previously~\cite{Haraguchi2015,Haraguchi2016,Haraguchi2017}, we show that intermediate stoichiometries are essential to generating a homogeneous QSL. Our experimental results agree well with the theoretical framework developed by Chen \emph{et al.}~\cite{Chen2016} and highlight a valuable new system for the study of QSL physics.

Polycrystalline samples of Li$_2$In$_{1-x}$Sc$_x$Mo$_3$O$_8$ were synthesized by solid-state reaction. A stoichiometric mixture of Li$_2$MoO$_4$, Sc$_2$O$_3$, In$_2$O$_3$, MoO$_3$, and Mo were ground together and pressed into 6 mm diameter, 60 mm long rods under 400 atm hydrostatic pressure which were placed in alumina crucibles and sealed in silica tubes at a pressure of $10^{-4}$ mbar. Finally, the samples were annealed for 48 hours at 850 C. Powder X-ray diffraction (XRD) patterns were recorded at room temperature with a HUBER Imaging Plate Guinier Camera 670 with Ge monochromatized Cu K $\alpha$1 radiation (1.54059 \AA). Mo-Mo bond lengths were refined by the Rietveld method~\cite{FullProf} with $\chi^2$ in the range 1-2 for all samples. Susceptibility measurements were performed at 2 T and specific heat measurements were carried out in zero field, with Quantum Design MPMS and PPMS systems. $\mu$SR measurements were carried out at TRIUMF in zero field (ZF) and longitudinal field (LF). Measurements in the range from 25~mK up to 3~K were performed with the samples affixed to an Ag cold finger of a dilution refrigerator. Higher temperature measurements were carried out in veto mode to eliminate the background asymmetry and were used to correct for the background present at low temperatures.

XRD spectra~\cite{SuppMat} reveal that as the In ions are replaced by smaller Sc ions, the lattice parameters decrease and, as seen in Fig.~\ref{xrd}(c), the ratio $a/c$ varies monotonically with a total change of about 1.4\%. It is important to investigate the evolution of the breathing parameter with $x$ and the XRD measurements reveal a non-monotonic behavior of $\lambda(x)$, as can be seen in Fig.~\ref{xrd}(d). The parent compound ($x = 0$) has the highest average degree of asymmetry, whereas at a concentration of 60\% In and 40\% Sc ($x = 0.6$) the lowest degree of asymmetry is attained. Meanwhile, the reported structure of LZMO~\cite{Sheckelton2012} corresponds to a breathing parameter of $\lambda \simeq 1.23$, making it closer to an ideal kagome lattice than the most symmetric sample in the series studied here.

\begin{figure}
\begin{center}
\includegraphics[width=3.35in]{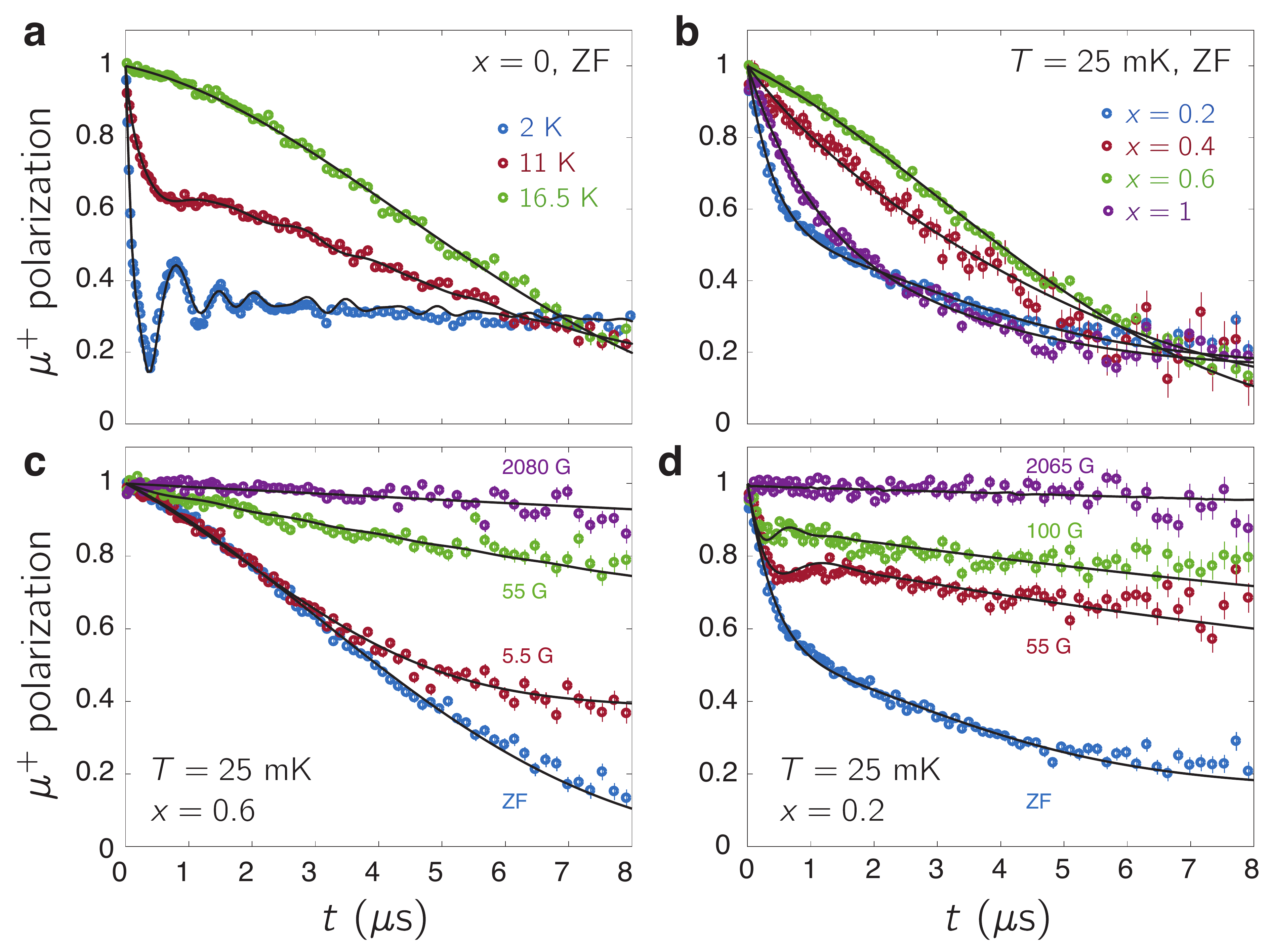}
\caption{ (a) Zero-field muon spin polarization $P(t)$ for Li$_2$InMo$_3$O$_8$ ($x = 0$). (b) Zero-field $P(t)$ measured at 25 mK for LiIn$_{1-x}$Sc$_x$Mo$_3$O$_8$ for different values of $x$. Polarization in various longitudinal fields for (c) $x=0.6$ and (d) $x=0.2$. The black lines are fits as described in the text. } \label{muSR}
\end{center}
\end{figure}

In general, the $\mu$SR polarization measured for our samples shows that the muon spins are influenced by a mix of fluctuating and static electron spins and the data are fitted with a two-component polarization function, $P_{\mathrm{tot}} = fP_S(t) + (1-f)P_D(t)$, where $P_S(t)$ is the polarization for the fraction $f$ of muons stopping in a static fraction (ordered or frozen) and $P_D(t)$ is the contribution from regions with dynamic electron spins, either QSL or paramagnetic phases. For the dynamic fraction, $P_D(t) = P_N(t)e^{-t/T_1}$ where $P_N(t)$ is a nuclear Gaussian Kubo-Toyabe function and $1/T_1$ is the spin-lattice relaxation rate. 

The ZF $\mu$SR asymmetry measured at 1.9~K  for Li$_2$InMo$_3$O$_8$ ($x  = 0$) shown in Fig.~\ref{muSR}(a) features a slowly decaying oscillation, demonstrating LRO with well defined internal fields consistent with NMR measurements of the same stoichiometry~\cite{Haraguchi2015}. $P_S(t)$ for this sample has thus been fitted to the static Lorentzian Koptev-Tarasov polarization function. Four distinct frequencies (1.1, 1.4, 2.0 and 3.3 MHz) are extracted, which correspond well to the four inequivalent oxygen sites. Select polarization curves at different temperatures in Fig.~\ref{muSR}(a) show the reduction of the oscillation frequencies (and order parameter) and the appearance of a dynamic fraction of the sample as the temperature is raised. The smallness of the observed frequencies is consistent with each spin-1/2 moment being highly distributed over a Mo$_3$O$_{13}$ cluster, similar to observations in systems of mixed-valence Ru dimers~\cite{Ziat2017}.

For $x=0.2$, $x=0.4$ and $x=1$, we find an inhomogeneous mix of disordered static magnetism (giving a quickly relaxing signal) and a weakly relaxing dynamic fraction as shown in Fig.~\ref{muSR}(b). The frozen fraction represents 49\%, 25\% and 43\% of these samples, respectively. On the other hand, $P(t)$ for $x = 0.6$ shows no indication of static fields originating from electron spins to as low as 25 mK, which suggests that the entire sample is in a homogeneous QSL phase. In fact, the $\mu$SR asymmetry profile for $x =0.6$ is very similar to that of LZMO~\cite{Sheckelton2014}.

\begin{figure}
\begin{center}
\includegraphics[width=3.35in]{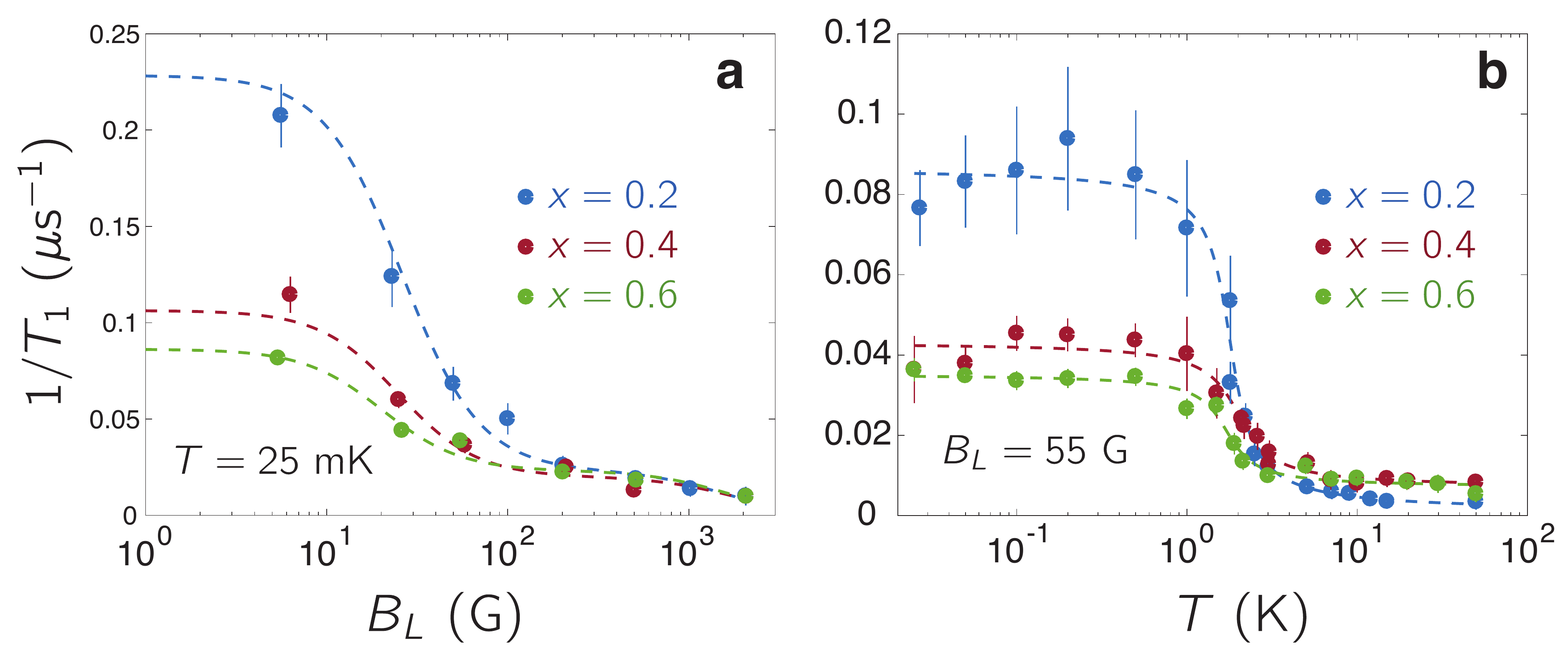}
\caption{ (a) Spin-lattice relaxation rate vs. longitudinal field at base temperature for the liquid phase of several samples, with fits given by Redfield theory with two different fluctuation frequencies. (b) Relaxation rate as a function of temperature in longitudinal field of 55 G, showing relaxation plateaus typical of QSL materials. Curves are guides to the eye.} \label{LambdaVsTandB}
\end{center}
\end{figure}

To fit the inhomogeneous samples, a Lorentzian Kubo-Toyabe function was used for $P_S(t)$. This fitting has been performed in zero- and longitudinal-field, $B_L$, as shown in Fig.~\ref{muSR}(d) and in supplemental material~\cite{SuppMat}. This analysis conclusively demonstrates that we have correctly identified the frozen and dynamic fractions of the sample since the muon spins are much more quickly decoupled from static than dynamic magnetism. For the homogeneous QSL sample, small $B_L$ quickly decouples the muon spins from the nuclear moments, but at higher field relaxation persists, indicating relaxation that is purely of dynamic origin, as seen in Fig.~\ref{muSR}(c). As shown in Fig.~\ref{LambdaVsTandB}(a), $1/T_1(B_L)$ for the QSL fractions is fairly well fit with Redfield theory~\cite{Slichter} using a sum of two characteristic fluctuation frequencies. Meanwhile $1/T_1$ of the liquid fractions shows relaxation plateaus in temperature below $\sim 1$ K, a common but still poorly understood feature of QSL candidates~\cite{Kermarrec2011,Quilliam2012bcso,Quilliam2016,Mendels2007}.

\begin{figure*}
\begin{center}
\includegraphics[width=7in,keepaspectratio=true]{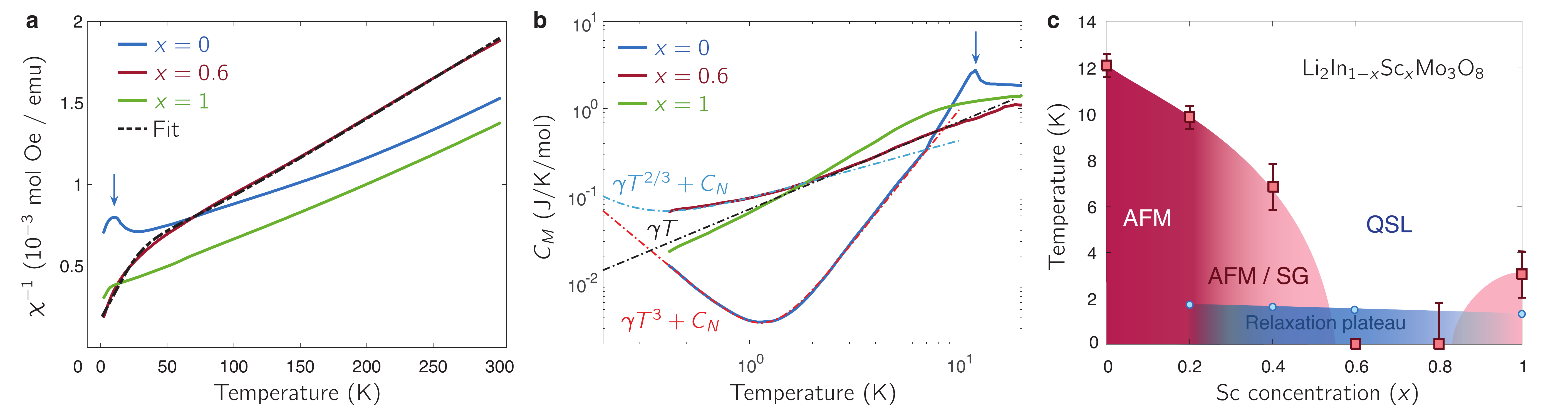}\\
\caption{(a) Temperature dependent inverse magnetic susceptibility $\chi^{-1}$ for select samples. For $x = 0$ a sharp feature at the onset of AFM order is indicated by an arrow at 11 K. $\chi^{-1}(T)$ for the homogeneous QSL sample, $x = 0.6$, shows two apparent Curie-Weiss regimes. The fit is described in the text. (b) The magnetic specific heat $C_M$ of select samples. The fit to the $x=0$ data is a $T^3$ power law plus a $C_N \propto T^{-2}$ nuclear contribution.  The specific heat of $x=0.6$ is compared with a $T^{2/3}$ power law plus nuclear contribution, as well as a $T$-linear dependence. (c) Magnetic phase diagram for Li$_2$In$_{1-x}$Sc$_x$Mo$_3$O$_8$. Red squares show the onset of freezing determined by specific heat (for $x=$0 and 0.2) $\mu$SR (for the remaining samples). The dark red region shows antiferromagnetic (AFM) ordering, whereas pink regions show spin freezing, either spin glass (SG) or disordered antiferromagnetism. The blue region shows the approximate temperature onset of the relaxation plateau in $\mu$SR.}  \label{Thermo}
\end{center}
\end{figure*}

Evidently the concentration of Sc does not monotonically change the ratio of static and QSL fraction, but rather there is an optimal concentration of $x = 0.6$ where a homogeneous QSL is stabilized. The phase diagram as a function of $x$, presented in Fig.~\ref{Thermo}(c), is highly correlated with the behavior of the breathing parameter, $\lambda(x)$, as shown in Fig.~\ref{xrd}(d). This suggests that the magnetic phenomenology of this material is intimately connected to the symmetry of the BKL and that past a critical value of $\lambda$ the system passes from antiferromagnetic to QSL. At critical values of $\lambda$, such as for $x=0.2$ and $x=1$, inhomogeneous phases result.

The way in which $\lambda$ influences the charge degrees of freedom, and consequently the spins, may be better understood with the magnetic susceptibility, $\chi$, measurements in Fig.~\ref{Thermo}(a). Our measurements of the end points of the series ($x=0$ and $x=1$) are consistent with previous work~\cite{Haraguchi2015}.  For intermediate concentrations $\chi(T)$ is very different. For the homogeneous QSL sample, $x=0.6$, $\chi^{-1}(T)$ displays two apparent linear Curie-Weiss regimes distinguished by different Curie constants and a smooth crossover between the two regimes. The $x=0.4$ and $x=0.8$ samples show similar behavior~\cite{SuppMat}. This strong, qualitative change in $\chi(T)$, even at high temperatures, implies that the effect of $x$ on the magnetic ground state is not simply an effect of disorder.

The temperature dependence of the susceptibility has been a central focus of the discussion surrounding the Mo$_3$O$_{13}$ cluster magnet family. Sheckelton \emph{et al.} first reported two Curie-Weiss regimes for the compound LZMO, where the Curie constant reduces to 1/3 of the high temperature value below a crossover at 96~K~\cite{Sheckelton2012}. They attributed this to the condensation of two-thirds of the spins into singlets~\cite{Sheckelton2012,Sheckelton2014}. Chen \emph{et al.}~\cite{Chen2016} proposed an alternative theory for the ``1/3-anomaly'' in $\chi^{-1}(T)$ whereby the low temperature regime corresponds to plaquette charge order (PCO). The PCO reconstructs the spinon bands with the lowest band splitting into 3 sub-bands. The lowest sub-band is completely filled with 2/3 of the spinons, becoming magnetically inert. The upper sub-band is partially filled with the remaining 1/3 spinons and these spinons contribute to $\chi$. Chen \emph{et al.}~\cite{Chen2014, Chen2016} argue that at the crossover temperature, PCO is destroyed and the full spin degrees of freedom are recovered. However, a transition between these two phases involves a spontaneous breaking of symmetry and should normally give rise to sharp thermodynamic features, the absence of which has been attributed to disorder~\cite{Chen2016}.

We propose an alternative mechanism for 1/3-anomaly. If the compounds $x=0.6$ and LiZn$_2$Mo$_3$O$_8$ are in the strong PCO regime, the energy scale required to break the PCO ($E_\mathrm{PCO} \sim t_1^3/V_2^2$) ought to be significantly larger than the energy gap, $\Delta E$, between filled and partially filled spinon sub-bands (which is governed by the next-nearest-neighbor interaction strength, $J'$), allowing for thermal excitation of spinons across the spinon gap while preserving PCO~\cite{Chen2018}. From a local perspective, each resonating hexagon in the PCO phase is composed of three coupled spins with a $S_\mathrm{tot} = 1/2$ ground state manifold and a $S_\mathrm{tot} = 3/2$ excited state. The magnetic susceptibility for non-interacting hexagons can be written as
\begin{equation}\label{ChiNonInt}
\chi_0 = \frac{\mu_0N_Ag^2\mu_{\beta}^2}{4k_B T} \frac{1+5e^{-\Delta E/k_B T}}{1+e^{-\Delta E/k_B T}} = \beta (T)\frac{C_0}{T}.
\end{equation}
The $S_\mathrm{tot} = 1/2$ ground state is doubly degenerate due to a pseudospin that represents the spatial configuration of entanglement in the resonating hexagon~\cite{Chen2016}. In a mean-field approximation, the interacting susceptibility then gives $\chi =  \beta (T) C_0/[T-\ \beta (T)\theta_W]$, naturally leading to two Curie-Weiss regimes with a ratio of 1/3 between the effective Curie constants $C_\mathrm{eff}= \beta (T) C_0$.

Eq.~\ref{ChiNonInt} gives an excellent fit of $\chi^{-1}(T)$ measured for sample $x=0.6$, shown in Fig.~\ref{Thermo}(a), where the parameters extracted from the fit are $C_0 = 0.264 \pm 0.001$ emu K Oe$^{-1}$ mol$^{-1}$, $\Delta E/k_B = 109 \pm 1$ K, and $ \theta_W = -46.3 \pm 0.5$ K. A fit of Eq.6 to the susceptibility data reported for LiZn$_2$Mo$_3$O$_8$~\cite{Sheckelton2012} is also successful (see supplemental material~\cite{SuppMat}), with fitting parameters $C_0 = 0.277 \pm 0.002$ emu K Oe$^{-1}$mol$^{-1}$, $\Delta E/k_B = 300 \pm 20$ K, and $\theta_W = -20 \pm 10$K. The same analysis can also be applied to other samples that are primarily spin liquids ($x=0.4$ and $x=0.8$) giving slightly different energy gaps. 

The magnetic specific heat, after lattice subtraction, for select samples is displayed in Fig.~\ref{Thermo}(b). As expected for LRO, the $x=0$ sample displays a peak at $T_N \simeq 12$~K and the appropriate power law, $C_M \propto T^3$, for gapless magnons. Below 1~K, the specific heat turns upward with a $T^{-2}$ power law which we attribute to the upper limit of a nuclear Schottky anomaly, $C_N$, likely originating from the $^{95}$Mo and $^{97}$Mo hyperfine couplings since the quadrupolar energy of $^{115}$In is not large enough~\cite{Haraguchi2017}. 

For samples that are primarily or entirely QSL ($x=$0.4, 0.6 and 0.8), there is no sharp peak and the $C_M(T)$ is much shallower. Between 1 and 10 K, $C_M\propto T$, but below 1 K $C_M$ becomes even shallower than linear.  This shallow temperature dependence of the specific heat in the order-free phase of this series of materials lends further evidence for a $U(1)$ QSL as predicted ~\cite{Chen2016,Motrunich2005,Lee2005}. It can be seen in Fig.~\ref{Thermo}(b) that if we apply the same nuclear contribution to the specific heat for the $x=0.6$ sample as was determined for the $x=0$ sample, a $T^{2/3}$ power law provides a reasonable fit to the data below $\sim 2$ K. Hence it is tempting to propose that this intermediate concentration has a $U(1)$ spin liquid state, similar to what has been proposed for the triangular organic QSLs~\cite{Motrunich2005,Pratt2011,Yamashita2008, Yamashita2009, Yamashita2011, Itou2008}, although there $C_M\propto T$ is observed~\cite{Yamashita2008,Yamashita2011}. For $x=1$ a somewhat steeper, $C_M\sim T^{1.4}$, is observed similar to the $T^{1.5}$ power law obtained in Ref.~\cite{Haraguchi2017}. The mixture of QSL and magnetic freezing may lead to an intermediate temperature dependence.

In conclusion, we have demonstrated a high degree of tunability of the series Li$_2$In$_{1-x}$Sc$_x$Mo$_3$O$_8$, through isovalant substitution of In with Sc. The magnetic phase diagram, Fig.~\ref{Thermo}(c), shows a strong correlation with the breathing parameter, with a homogeneous QSL phase in the most symmetric sample at $x=0.6$, suggesting that $\lambda$ is the principal controlling parameter. The nature of $\chi(T)$ also varies substantially with $x$. Notably, in the range of $0.4 < x <0.8$, $\chi^{-1}(T)$ is very similar to that of the QSL LZMO, with two apparent Curie-Weiss regimes. This observation fits well with the theory of Chen \emph{et al.}~\cite{Chen2016} predicting the 1/3-anomaly in the PCO phase, which should be stabilized by small $\lambda$. We propose that the 1/3-anomaly originates from thermal excitations of the resonating hexagons from the $S_\mathrm{tot} = 1/2$ ground state to a $S_\mathrm{tot} = 3/2$ excited state. Since smaller $\lambda$ and the 1/3-anomaly seem to be associated with a QSL ground state, the spins in the PCO phase appear to be more frustrated than in the Type-I CMI. Indeed the specific heat of the homogeneous QSL at $x=0.6$ has a particularly shallow temperature dependence, possibly consistent with a $U(1)$ QSL~\cite{Motrunich2005,Lee2005}. 

This work has therefore provided a likely resolution to the debate surrounding LZMO~\cite{Sheckelton2012}. An alternative scenario to explain the 1/3-anomaly in LZMO has been put forward by Flint and Lee~\cite{Flint2013}, wherein the electrons are localized on the up-triangles but two thirds of the clusters rotate, generating an emergent honeycomb lattice, thereby leaving 1/3 of the spins as weakly connected ``orphan'' spins.  However, we find no natural reason that changes in $\lambda$ would encourage rotation of Mo$_3$O$_{13}$ clusters and the 1/3 of the spins that remain active at low temperature exhibit a strongly negative Curie-Weiss constant, $\Theta_W \simeq -46$ K, meaning they cannot be described as orphan spins. 

Valuable future work on this series could include direct measurements of charge order with resonant X-ray spectroscopy, although the changes in local charge density will be rather small, as well as a search for thermodynamic indications of charge-ordering at higher temperatures. Furthermore, it would be interesting to study the parent compounds under applied pressure instead of chemical pressure, potentially tuning the system into a QSL phase without introducing structural disorder. Indeed the role of disorder in either destabilizing or even generating QSL-like phases remains a contentious issue in the field~\cite{Kawamura2014}.  Furthermore, although the model proposed by Chen \emph{et al.} ~\cite{Chen2016} is consistent with our observations, many assumptions have been made regarding the appropriate Hamiltonian for these materials which should be validated with detailed electronic structure calculations.

\begin{acknowledgements}

We are grateful to the staff of the Centre for Molecular and Materials Science at TRIUMF for extensive technical support, in particular G. Morris, B. Hitti,  D. Arseneau, and I. MacKenzie. We also acknowledge helpful conversations with Y. B. Kim, G. Chen, H.-Y. Kee, M. Gingras, A.-M. Tremblay, F. Bert and P. Mendels.  A. A.-S. and J. Q. acknowledge funding through NSERC, FRQNT, CFI and CFREF grants. H.~D.~Z. acknowledges support from the Ministry of Science and Technology of China with grant number 2016YFA0300500. R.~S. and H.~D.~Z. acknowledge support from NSF-DMR with grant number NSF-DMR-1350002. X.~F.~S. acknowledges support from the National Natural Science Foundation of China (Grant Nos. 11374277, U1532147) and the National Basic Research Program of China (Grant Nos. 2015CB921201, 2016YFA0300103).

\end{acknowledgements}


%

\end{document}